\documentclass[sigconf]{acmart}

\usepackage{amsmath,amsfonts}
\usepackage{algorithm}
\usepackage{algpseudocode} 
\usepackage{graphicx}
\usepackage{textcomp}
\usepackage{xcolor}
\usepackage{eucal}
\usepackage[T1]{fontenc}

\usepackage{subcaption}
\usepackage{setspace}
\usepackage{rotating}

\usepackage{hyperref}
\hypersetup{breaklinks=true}

\usepackage{enumitem}
\setlist[itemize]{noitemsep, topsep=0pt}

\AtBeginDocument{%
  \providecommand\BibTeX{{%
    \normalfont B\kern-0.5em{\scshape i\kern-0.25em b}\kern-0.8em\TeX}}}

\setcopyright{acmcopyright}
\copyrightyear{2020}
\acmYear{2020}
\acmDOI{10.1145/1122445.1122456}

\acmConference{}
\acmBooktitle{}
\acmPrice{15.00}
\acmISBN{978-1-4503-XXXX-X/18/06}

\begin{document}

\title{DOOM: A Novel Adversarial-DRL-based Op-Code Level Metamorphic Malware Obfuscator for the enhancement of IDS}

\author{Mohit Sewak}
\affiliation{%
  \institution{Department of CS \& IS, Goa Campus}
  \city{BITS Pilani, Goa, India.} 
}
\email{p20150023@goa.bits-pilani.ac.in}
\author{Sanjay K. Sahay}
\affiliation{
  \institution{Department of CS \& IS, Goa Campus,}
  \city{BITS Pilani, Goa, India.} 
}
\email{ssahay@goa.bits-pilani.ac.in}
\author{Hemant Rathore}
\affiliation{
  \institution{Department of CS \& IS, Goa Campus}
  \city{BITS Pilani, Goa, India.} 
}
\email{hemantr@goa.bits-pilani.ac.in}

\renewcommand{\shortauthors}{Sewak, et al.}

\begin{abstract}
We designed and developed DOOM (Adversarial-\textbf{D}RL based \textbf{O}p-code level \textbf{O}bfuscator to generate \textbf{M}etamorphic malware), a novel system that uses \textbf{adversarial deep reinforcement learning} to obfuscate malware at the op-code level for the enhancement of IDS. The ultimate goal of DOOM is not to give a potent weapon in the hands of cyber-attackers, but to create defensive-mechanisms against advanced zero-day attacks. 
Experimental results indicate that the obfuscated malware created by DOOM could effectively mimic multiple-simultaneous \textit{zero-day} attacks. To the best of our knowledge, DOOM is the first system that could generate obfuscated malware detailed to individual op-code level. DOOM is also the first-ever system to use efficient continuous action control based deep reinforcement learning in the area of malware generation and defense. Experimental results indicate that over \textbf{$\mathbf{67\%}$} of the metamorphic malware generated by DOOM could easily evade detection from even the most potent IDS. This achievement gains significance, as with this, even IDS augment with advanced routing sub-system can be easily evaded by the malware generated by DOOM.
\end{abstract}



\maketitle

\section{Introduction}\label{sec:introduction}
We present a novel system to create obfuscated/ metamorphic malware using \textit{Adversarial Deep Reinforcement Learning} (A-DRL), and named it as \textbf{\textit{DOOM}} (Adversarial- \textbf{D}RL based \textbf{O}p-code level \textbf{O}bfuscator to generate \textbf{M}etamorphic). Unlike some existing systems, DOOM can even create obfuscations of malware to their respective op-code-sequence level. Doing so, the output of DOOM mimics an actual metamorphic malware at op-code level. Therefore, an existing \textit{Intrusion Detection Systems} (IDS) is very much likely to encounter such a metamorphic malware \cite{sahay2020evolution} created by DOOM in real-life than a random perturbations of an binary that does not resemble a malware or even preserve the actual file functionality. 
\par
Our op-code level obfuscations generated by the DOOM can be used for:
\begin{itemize}
    \item Improving the IDS's classifier against new or metamorphic variants of existing malware.
    \item Training/ augmenting other \textit{internal} sub-systems of the IDS with the capability to de-obfuscate the incoming file's features vector before sending it to the IDS's classifier.
    \item Creating/ training other \textit{external} sub-systems for \textbf{\textit{normalizing}} obfuscations of different variant of existing malware. This can augment any existing IDS with metamorphic detection capabilities without warranting any changes.
\end{itemize}


\par The instruction-set for any processing architecture may contain thousands of op-codes. Therefore the \textit{Markov Decision Process} (MDP) that can mimic the obfuscation of a malware holistically by inserting number of junk \textit{instructions} in different sequence-combination and consequently represents a very high-cardinality action-space. Solving such an MDP require a Reinforcement Learning (RL) agents that could find efficient solution under these constraints. However, training an RL agent for the given purpose is not a trivial task. Hence in DOOM we use Proximal Policy Optimization (PPO) \cite{PPO} algorithm-based agents in an adversarial setup to solve the MDP to create the desired obfuscations. Experimental results indicate that the metamorphic malware created by DOOM could evade (classified as non-malicious or ambiguous) even the most potent malware-IDS \cite{SewakSNPD} and hence could mimic extreme multiple-simultaneous \textit{zero-day} attacks from metamorphic malware. To the best of our knowledge, DOOM is the first-ever system that takes an MDP-based approach for generating \textit{sample-efficient} \textit{op-code-level} obfuscation.
The system use adversarial learning mechanism, in which (multiple) PPO algorithm-based DRL agents acts as  \textit{Generator} $\mathcal{G}$ network(s) in the adversarial-setup, and the IDS acts as the \textit{Discriminator} $\mathcal{D}$ network. In our experiments, we use the complete IDS (including pre-processing, feature-selection and transformation, and classifier sub-components) that claimed the best performance (accuracy of $99.21\%$ with an FPR of $0.19$) \cite{SewakSNPD} with a benchmark malware dataset \cite{Malicia}.

\section{Related Work} \label{sec:related-work}
There has been many attempts to generate obfuscations at the code level \cite{metamorphic_obfuscation_Borello2008}, but these are not scalable. Later efforts were also made to use machine-learning (ML) models \cite{metamorphic_virus_generator_desai} to automate the obfuscation mechanism. However, these ML methods does not effectively replicate the advanced metamorphic attack  required to train an adversarial mechanism. There has been attempt to use Convolutional Neural Networks (CNN) based Generative Adversarial Networks (GANs) \cite{adversarial-gan},\cite{MalGAN},\cite{IDSGAN}, \cite{UsamaGAN} as well. But these Deep Learning (DL) \cite{sewak2020overview} systems mostly use supervised training approaches with specific algorithms and the adversarial-learning produced from such mechanisms is immune to secondary gradient-attack. Recently DRL \cite{Sewak-DRL}, especially Q Learning has been utilized \cite{drl-evading-botnet} to alter the \textit{binary-code} of the file to evade attacks. Systems based on creating perturbations at binary-code level are not only limited to very small action-space of MDP (limited to adding some specific 4-bit code in \cite{drl-evading-botnet}), and also can not be used to mimic an actual malware, because it require a code/op-code level obfuscation.
To the best of our knowledge, DOOM is the first-ever system to work directly on the numerous of op-code-sequence combinations (hence requires solving a large-action space reinforcement problem) to generate obfuscated feature vectors that could be used to train any adversarial-defense solution (RL or DRL based) that works on op-code frequency features for the malware detection.

\section{DOOM's Architecture}\label{sec:process-flow}
\begin{figure*}
    \centering
    \includegraphics[width=\linewidth, height=2in]{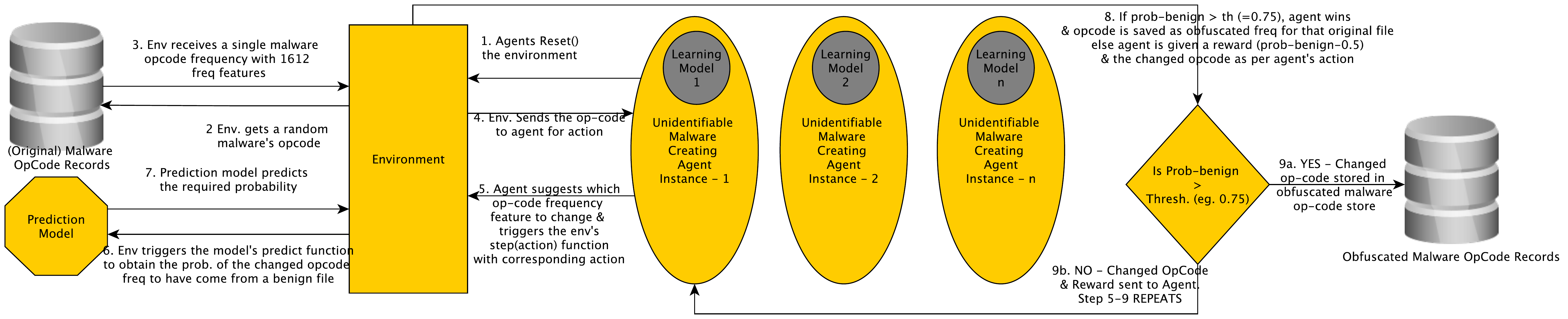}
    \caption{Design of the DOOM system}
    \label{fig:process_flow}
\end{figure*}

As shown in figure \ref{fig:process_flow}, the architecture of the \textit{DOOM} broadly consists of four subsystems, which are
\begin{enumerate}
    \item The op-code repository of the original-malware feature-vectors and its subsequent obfuscated instances generated by DOOM.
    \item Repository of existing trained IDS to act as the adversary (Discriminator $\mathcal{D}$ Network as in GAN).
    \item A custom RL-environment (to emulate the MDP with which the agents could interact and learn against).
    \item The Obfuscating DRL Agent(s).
\end{enumerate}

\par The DRL agents interact with the environment to train on \textit{episodic} tasks. In each step of an episode, the agent alters the op-code frequency-vector of a candidate malware as presented by the environment. The agent may choose to insert op-codes, even the ones which though present in the instruction set of the architecture but not present in the file. 
During the initialization of an \textit{episode}, the environment fetches a randomly selected malware's op-code frequency feature vector from the malware repository. 
Then the environment scores the obtained feature vector using the available IDS. The IDS acts as an adversary and provides the (initial) malicious (prediction) probability ($\mathbb{P_\text{non-malicious}}$) of the selected malware. This initial malicious probability is stored in a metadata which is later used to compute the increase/decrease in $\mathbb{P_\text{non-malicious}}$ after each episode by the specific agent.

\textit{Multiple} such DRL agents can be instantiated and trained with varying degree of dissimilarity from the other DRL agents so that each could learn to obfuscate a given malware using a slightly different action-policy and subsequently producing multiple dissimilar metamorphic instances of the same malware variant.

\subsection{Ethically-Safe Mechanism} \label{sec:safety-mechanism}
There are many malware creation tool-kits available in public domain. Such tool-kits do not use advanced artificial intelligence to create virtually undetectable obfuscations, like the ones that DOOM can generate. Therefore a system like DOOM, in wrong hands could have serious implications. Therefore to obviate such negative outcomes we have purposely designed a process that ensures that the obfuscation component of the system would work only at the op-code level and could not be used to create a malicious \textit{executable} file. 

\subsection{Functionality-Preserving Metamorphism} \label{sec:functionality-preserving}
DOOM detects the malware at the instruction-level and also create obfuscations at the instruction-level. The functionality of a given program is represented by the sequence of the instructions available in its assembly produced by the compiler. As described in section \ref{sec:process-flow}, DOOM only inserts junk instructions and does not remove the existing ones. Doing so, DOOM preserves the intended functionality of the program. But with such capabilities risks arise that DOOM could be used to create an actual metamorphic malware. Therefore owing to the desired outcome of functionality-preservation, we had ensure the safety mechanism as described in section \ref{sec:safety-mechanism}, which ensures that despite of all the theoretical purposes (for the claims and experiments) the functionality is preserved, in practice generating a malicious PE remains \textit{intractable}.

\section{DRL Agent used}\label{sec:rl-agent}

Some of the most popular DRL agents, like the \textit{Deep Q Networks} (DQN) \cite{DQN_Nature}, \textit{Double DQN}  \cite{Double_DQN} perform poorly for large/ continuous action-space of MDP. Deterministic Policy Gradient \cite{DPG} based DRL approaches like the \textit{Deep Deterministic Policy Gradient} \cite{DDPG} claims to deliver the best-in-class performance on MDPs involving large, and even in continuous action-space. Deterministic Policy Gradient is a simplification of the Stochastic Policy Gradient and can be given as 
\begin{equation}
  \nabla_\theta (J_\theta) =  \mathbb{E}_{\tau \sim \pi_\theta (\tau)} [\nabla_\theta \log \pi_\theta(\tau) r(\tau)]  
  \label{eq:policy-gradient}
\end{equation}

\noindent Here, $J$ is the policy-utility function with parameter-vector $\theta$, that gives the expectancy of rewards ($r$) over a trajectory $\tau$ while following a policy $\pi$. Based on deterministic-policy-gradient method, Trust Region Policy Optimization (TRPO) algorithm  optimize the policy-gradient (equation \ref{eq:policy-gradient}) using linear-search approach. The problem with such algorithms like TRPO is that their line-search-based policy gradient update (used during optimization) either generates too big updates (for updates involving non-linear trajectory makes the update go beyond the target) or makes the learning too slow. This is where the \textit{Proximal Policy Optimization} \cite{PPO}, an enhanced variant of TRPO excels. PPO algorithm works similar to TRPO but is more computationally efficient, as it uses a linear-variant of the gradient-update called the \textit{Fisher Information matrix}. In TRPO, the trajectory is sampled from the policy as it existed in previous time-step $(\pi_{\theta} = Q(x))$, whereas the expectancy over such collected samples are used to update the policy at next time-step $(\pi_{\theta\prime} = P(x))$. When the ratio of expectancy over the two trajectory distributions vary significantly as in the case of linear gradient-update in PPO, assumption that $P(x)\approx Q(x)$ may not hold, leading to high variance in policy updates. To avoid this we use \textit{Objective Clipping.} In this mechanism if the probability ratio between the two trajectory's policies is not in the range $[(1-\epsilon),(1+\epsilon)]$ the estimated \textit{advantage} is clipped.

\section{Experiments and Results}\label{sec:results}
\begin{figure}[htb]
    \centering
    \includegraphics[width=\linewidth, height=2.0in]{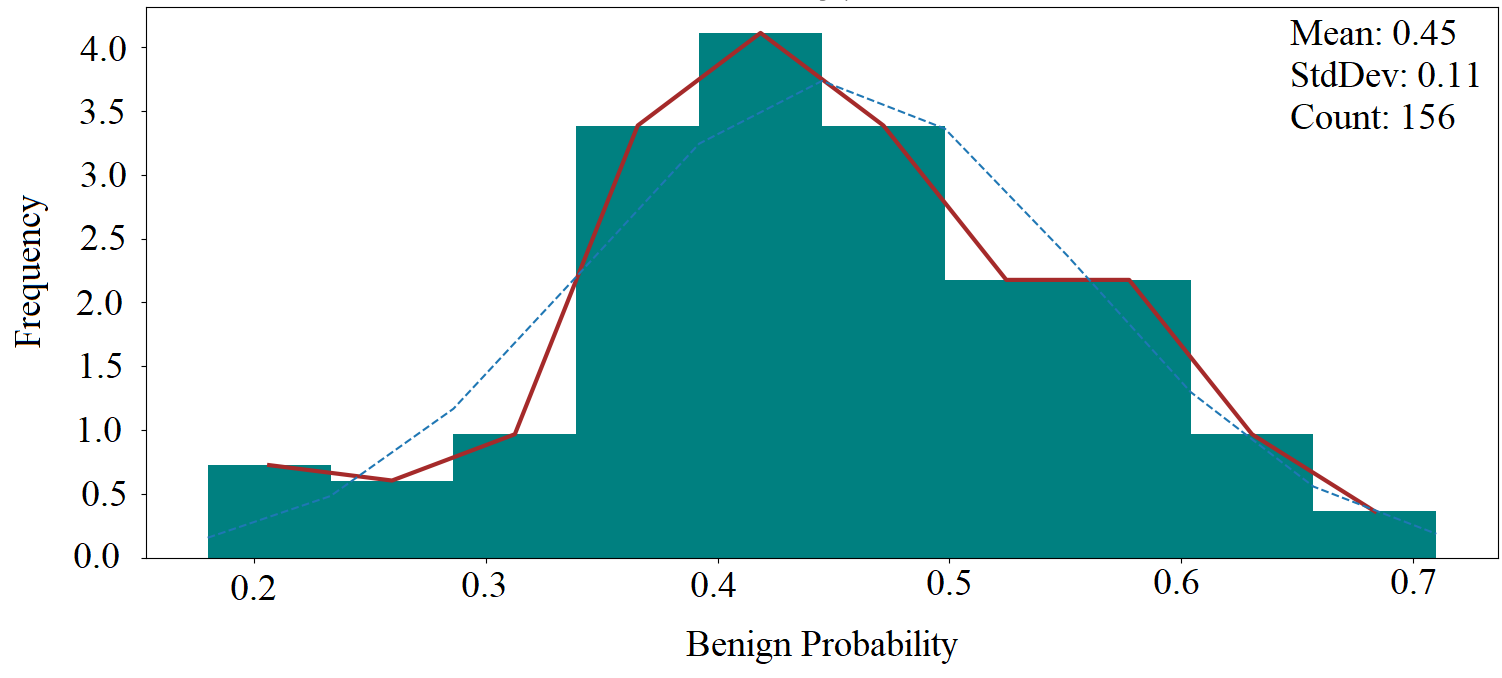}
    \caption{Histogram of uplift in miss-classification probability for a sample Agent}
    \label{fig:training_statistics_agent_1}
\end{figure}

\begin{figure}[htb]
    \centering
    \includegraphics[width=\linewidth, height=1.7in]{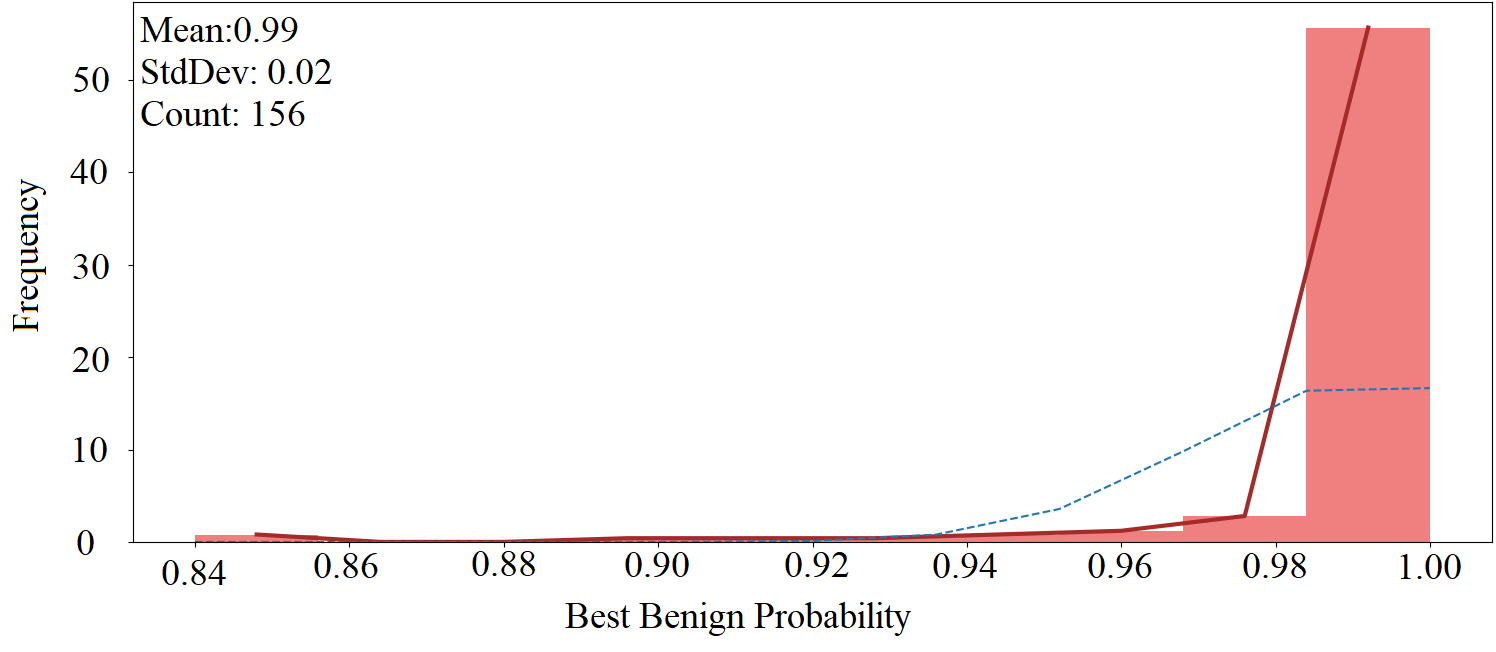}
    \caption{Histogram of op-code Similarity for a sample Agent}
    \label{fig:opcode_similarity_agent_1}
\end{figure}

Several versions of PPO agents were trained with different instances of the custom environment. The mean miss-classification probability of all the files against all the trained agents indicate that the resulting trained agents could obfuscate most of the malware and uplift their metamorphism ($\mathbb{P_{\text{non-malicious}}}$) (probability of miss-classification) to substantial degree to evade even the best IDS. Even the IDS which were trained with the original malware samples failed to detect their corresponding metamorphic instances.
As shown in figure \ref{fig:training_statistics_agent_1}, the mean probability of the non-detectable malware file ($\mathbb{P}_{NDMF}$) has been uplifted by $\ge 0.45$ (from almost 0.0) indicating that the IDS can not effectively detect such obfuscated instances of malware. 
Another interesting observation is related to the op-code similarity between the original malware variants and its generated obfuscated version. 
Figure \ref{fig:opcode_similarity_agent_1} shows a histogram of similarity of the generated op-code sequence to that of the original malware,
and indicates that the op-code frequency vector for the obfuscated variant is very similar to the original malware variant.




\section{Discussion}\label{sec:discussion}
The insights from these observations are very critical as it indicates that with minimal efforts (adding additional op-code instructions using methods like junk-code insertion etc.) new obfuscations with very high metamorphic probability can be generated by DOOM. Another important observation is that even with a net-additive obfuscation of the metamorphic malware the resulting feature-vector is very similar to that of the original. Hence even for advanced multi-classifier IDS, the same classifier is likely to score the metamorphism which may had failed to detect it earlier, thus further enhancing the evasion probability. 
This observation is important as many advanced IDS, may have an unsupervised routing sub-component that helps to improve the IDS performance by routing the candidate file to the appropriate classifier based on the unsupervised routing sub-component. This routing sub-component could use clustering \cite{hemant_clustering} to route the detection to different classifier based on the file's cluster-assignment. Owing to a high feature (and hence file-size) similarity, the resulting obfuscations from DOOM are much likely to be routed to the same classifier of an IDS, and hence the evasion probabilities as reported remains relevant for such advanced IDS as well.

\section{Conclusion}\label{sec:conclusion}
We designed and developed DOOM, a system to create obfuscations at op-code level to generate metamorphic malware. The obfuscations thus generated could be used to augment the capabilities of an IDS in multiple ways. It uses advanced PPO algorithm based DRL agents, and a custom RL environment and reward-functions. The experimental results shows that DOOM could effectively obfuscate and generate multiple metamorphic instances of malware and successfully achieve the primary objective ($\mathbb{E}[\mathbb{P}_{NDMF}] \approx 0.5$) for almost all of their metamorphic instances generated by it. Additionally, DOOM successfully achieved both the primary and secondary objectives for nearly one-third of the metamorphic obfuscations, and the generated malware were generated similar enough to the original variant, to ensure that they could evade even advanced IDS with multiple classifiers and sophisticated routing mechanism.
Hence, the metamorphic malware instances generated by DOOM can be used for designing and training defensive mechanisms that could avert even extreme multiple-simultaneous \textit{zero day} attack by advanced metamorphic malware. The ultimate goal of DOOM is not to give a potent weapon in the hands of cyber-attackers, but to create defensive-mechanisms against advanced zero-day attacks, and the work in this direction is in progress.

\bibliographystyle{ACM-Reference-Format}
\bibliography{main}

\end{document}